\documentclass[prd,12pt,aps,double-spaced,floatfix,showkeys]{revtex4-1}
\usepackage{graphicx,float,amsmath,bbm,bm,array,subfigure}
\usepackage{mathtools}
\usepackage[linktocpage=true,colorlinks=true,linkcolor=blue,citecolor=blue]{hyperref}

\begin{document}
	
	\title {Modified Excluded Volume Hadron Resonance Gas Model with Lorentz Contraction}
	\author{Somenath Pal$^1$}
	\email{somenathpal1@gmail.com}

	\author{Abhijit Bhattacharyya$^1$}
	\email{abhattacharyyacu@gmail.com}
	
	\author{Rajarshi Ray$^2$}
	\email{rajarshi@jcbose.ac.in}
	
	\affiliation{$^1$Department of Physics, University of Calcutta,\\ 92, A.P.C. Road, Kolkata-700009, India\\$^2$Center for Astroparticle Physics \&
		Space Science, \\ Block-EN, Sector-V, Salt Lake, Kolkata 700091, India 
		\\ \& \\ 
		Department of Physics, Bose Institute,\\
		93/1, A. P. C Road, Kolkata 700009, India}
	
	\def\be{\begin{equation}}
		\def\ee{\end{equation}}
	\def\bearr{\begin{eqnarray}}
		\def\eearr{\end{eqnarray}}
	\def\zbf#1{{\bf {#1}}}
	\def\bfm#1{\mbox{\boldmath $#1$}}
	\def\hf{\frac{1}{2}}
	\def\sl{\hspace{-0.15cm}/}
	\def\omit#1{_{\!\rlap{$\scriptscriptstyle \backslash$}
			{\scriptscriptstyle #1}}}
	\def\vec#1{\mathchoice
		{\mbox{\boldmath $#1$}}
		{\mbox{\boldmath $#1$}}
		{\mbox{\boldmath $\scriptstyle #1$}}
		{\mbox{\boldmath $\scriptscriptstyle #1$}}
	}

\begin{abstract}
In this work we discuss a modified version of Excluded Volume Hadron
Resonance Gas model and also study the effect of Lorentz
contraction of the excluded volume on scaled pressure
and susceptibilities of conserved charges. 
 We find that the Lorentz contraction, coupled with the variety of excluded volume parameters
 reproduce the lattice QCD data quite satisfactorily. 

\end{abstract}

\maketitle

\section{Introduction}

Studies of strongly interacting matter at high temperatures and/or high
densities have been of great interest for some decades. In the early universe, a few microseconds after the Big Bang, strongly interacting matter is expected
to have existed in the colour charge deconfined quark-gluon phase~\cite{big_bang}. On the other hand, dense strongly interacting matter can be found inside neutron stars~\cite{neutron_star}. There are several ongoing and up-coming
experiments with ultra-relativistic heavy ion collision which are
recreating such phases of strongly interacting matter. Among the experimental
programmes, Large Hadron Collider (LHC) at CERN, Geneva and Relativistic Heavy Ion Collider (RHIC) at Brookhaven,
New York, have already enriched lot of our understanding in this
direction. The ongoing experiments in these facilities as well as the
upcoming facilities at GSI, Darmstadt and at JINR, Dubna, are expected to
further expand our knowledge about the various facets of strongly 
interacting matter at high temperature and density.

The experimental investigations have been very well supported by
significant advancements of the theoretical approaches.  One of the main
objectives of these explorations is to understand the thermodynamic
phases as well as the phase diagram of strong interactions at high
temperatures and high densities.  At high temperature and low density,
the phase boundary between hadronic and quark-gluon matter is found to be a
crossover~\cite{crossover1, crossover2}, while at low temperature and
high density there is possibly a first order phase transition~\cite{1st_order1, 1st_order2, 1st_order3, 1st_order4, 1st_order5,
1st_order6}.  Thereby the possibility of observing signatures of a
critical end point (CEP)~\cite{CEP1, CEP2, CEP3, CEP4} has become an
extremely active field of research.

Though quantum chromodynamics (QCD) is the theory of strong interactions, the traditional perturbative methods of field theory is inadequate because the strong
interaction coupling may not be small for the temperatures and densities
concerned.  Lattice QCD (LQCD) is the most important non-perturbative
tool that can describe strongly interacting matter at high
temperatures~\cite{crossover1,LQCD1, LQCD2, LQCD3, LQCD4, LQCD5, LQCD6,
LQCD7, LQCD8, LQCD9, LQCD10, LQCD11, LQCD12, LQCD13, LQCD14, LQCD15}.
However, the Monte Carlo techniques of LQCD cannot be applied to a system
with finite baryon chemical potential $\mu_B$, as the fermion
determinant becomes complex. However, the Taylor expansion of
thermodynamic quantities around $\mu_B=0$, for a given temperature $T$,
can be used until $\mu_B$ is close to a phase boundary. For this reason,
people build effective models to study properties of strongly
interacting matter in non-perturbative domain.  Some examples of such
models are Polyakov Loop Extended Nambu-Jona-Lasinio (PNJL) Model~\cite{PNJL1,PNJL2,PNJL4,
PNJL10,PNJL11,PNJL12,
PNJL15},
Hadron Resonance Gas (HRG) Model~\cite{HRG1,HRG2,HRG3,HRG4,HRG5,HRG6,HRG7,HRG8,HRG9,HRG10}, Polyakov-Quark-Meson (PQM) 
model~\cite{PQM1,PQM2,PQM3,PQM4}, Chiral Perturbation Theory~\cite{chiral1}, etc., which successfully describe some aspects of strongly interacting matter.

HRG model is used to describe a system of dilute gas consisting of hadrons and is based on Dashen, Ma, Bernstein theorem~\cite{dashen}. In this model attractive interactions among hadrons are taken care of by considering the unstable resonance particles as stable particles. HRG model successfully describes the experimental data of a system at freeze out~\cite{HRG_freezeout1}. However, repulsive interaction among the particles in the hadronic system is also important~\cite{EVHRG1}. This is taken into account in the modified version of HRG model, namely, Excluded Volume Hadron Resonance Gas (EVHRG) model where repulsive interaction comes into play due to finite excluded hardcore volume of the particles~\cite{EVHRG2,EVHRG3,EVHRG4,EVHRG5,EVHRG6,EVHRG7,EVHRG8}.

Recently, several works have been done on EVHRG using different sizes for different hadrons. 
In Ref.~\cite{alba}, the authors have studied the effect of excluded volume on the equation of state. In particular, they have looked in to the pressure and the trace 
of the energy momentum tensor and compared their results with the lattice data. They have found that the best fits are obtained when the excluded volume is inversely 
proportional to the mass of the particle. Recently there is a renewed interest in studying multiplicity data~\cite{diptak1,diptak2}. However, a more prudent approach has been followed in Ref.~\cite{gupta} 
where HRG model has been studied in the light of both multiplicity and fluctuation data. The authors have treated temperature and chemical potential as parameters 
and tried to fit those by fitting the data using non-interacting HRG model. Fluctuations of baryon number and strangeness within HRG model with repulsive mean field approach with the effect of missing resonances has been studied in Ref.~\cite{Huovinen:2018ziu}. HRG model with parity-doubled baryons having temperature dependent mass was used to calculate charge susceptibilities in Ref.~\cite{Morita:2017hgr} and it was found that mass reduction at high temperatures overshoots the lattice data. The contribution of heavy resonances through exponential Hagedorn mass spectrum to fluctuations of conserved charges was discussed in Ref.~\cite{Lo:2015cca}. In Ref.~\cite{Lo:2017ldt}, a comparison between EVHRG hardcore repulsion and interaction based on S-matrix in HRG framework was considered and it was found that for a $\pi N \Delta$ system, there is good agreement between the two approaches with excluded volume radius $R = 0.3 fm$ for pions and nucleons. In Ref.~\cite{Bugaev:2012wp}, the authors have shown that the mid-rapidity data for hadron yield ratios at AGS, SPS and two highest RHIC energies, are best fit when $R_\pi = 0, R_K = 0.35 fm, R_{mesons} = 0.35 fm, R_{baryon} = 0.5 fm$. In Ref.~\cite{Vovchenko:2017zpj}, a choice of 2nd virial coefficient for nucleons = $3.42 fm^3$ was found to generate the ground state nuclear properties well. Some other choices can be found in Refs.~\cite{Samanta:2017yhh,Vovchenko:2018cnf,Vovchenko:2020lju}. This indicates that there is no strict consensus about the value of excluded volume of hadrons.

Fluctuations of conserved charges are useful indicators of phase
transition between hadronic and quark gluon plasma phase. Existence of
CEP can also be indicated by divergent fluctuations. One can calculate
charge susceptibilities which are related to fluctuations via
fluctuation-dissipation theorem. If net baryon number of the system is
small then transition from hadronic to QGP phase is continuous and
fluctuations are expected not to show singular behaviour. On the other
hand Lattice QCD calculation shows that at small chemical potentials,
susceptibilities show rapid increment near the crossover region. Higher
order susceptibilities are considered to be more sensitive to phase
transition. Fluctuation of conserved charges has been studied
in Refs.~\cite{fluctuation1, fluctuation2, fluctuation4,volume2,fuku2008,roessner1,friman,roessner2,fuku3,schaefer1,
schaefer2,schaefer3,schaefer4,Borsanyi:2018grb,bazavov:2020}.

This paper is organised as follows. In the next section we briefly
discuss the Hadron Resonance Gas (HRG) model and its interacting version
Excluded Volume Hadron Resonance Gas (EVHRG) model briefly. We also
introduce a modified version of EVHRG model, namely, MEVHRG model and its
extended version Lorentz contracted MEVHRG model or LMEVHRG model.
In section (III) we discuss our results of pressure and fluctuation  of conserved charges. 
We also compare our results with the recent LQCD data. 
 Finally, in section (IV) we conclude our findings.

\section{Hadron Resonance Gas Model}

Here we present a brief discussion of Hadron Resonance Gas (HRG) model
and its interacting version namely Excluded Volume Hadron Resonance Gas
(EVHRG) Model. More details about these models can be found in Refs.~\cite{HRG1,HRG2,HRG3,HRG4,HRG5,HRG6,HRG7,HRG8,HRG9,EVHRG1,EVHRG2,EVHRG3,EVHRG4,EVHRG5,EVHRG6,EVHRG7,EVHRG8}.

\subsection{Pure HRG} In HRG model, a dilute system of strongly
interacting matter is considered as a gas of free resonances. The
attractive interactions are taken care of by considering all the
resonances as stable particles~\cite{dashen}.

The grand canonical partition function is $\mathcal{Z}^{id}$.
\begin{equation}
\ln \mathcal{Z}^{id}=\sum_{i}\ln\mathcal{Z}^{id}_i
\end{equation}
Here the sum runs over all hadron species and $"id"$ in superscript refers to ideal gas.
\begin{equation}
\ln{\mathcal{Z}^{id}_i}=\pm\frac{Vg_i}{2\pi^2}\int_{0}^{\infty}p^2\ dp\ln\{1\pm exp[-(E_i-\mu_i)/T]\}
\end{equation}
Here $V$ is the volume of the system, 
$g_i$ is the degeneracy factor, 
$p$ is the momentum of a particle, 
$E_i=\sqrt{p^2+m^2_i}$ is the energy of a single particle, 
$m_i$ is the mass of particle species $i$, 
$T$ is the temperature, 
$\mu_i=B_i\mu_B+S_i\mu_S+Q_i\mu_Q$ is the chemical potential of particle species $i$, 
$B_i, Q_i,S_i$ are baryon number, electric charge and strangeness of $i$'th hadron species respectively and $\mu$'s are respective chemical potentials, 
+ sign is for fermions and - sign is for bosons.\\

From partition function we can calculate various thermodynamic quantities like pressure $P_i$, number density $n_i$ as follows
\begin{equation}
P_i^{id}=\frac{T}{V}\ln\mathcal{Z}_i^{id}=\pm\frac{Tg_i}{2\pi^2}\int_{0}^{\infty}p^2\ dp\ln\{1\pm \exp[-(E_i-\mu_i)/T]\}
\end{equation}
\begin{equation}
n_i^{id}=\frac{T}{V}\Big(\frac{\partial \ln \mathcal{Z}^{id}_i}{\partial \mu_i}\Big)_{V,T}=\frac{g_i}{2\pi^2}\int_{0}^{\infty}\frac{p^2\ dp}{\exp[(E_i-\mu_i)/T]\pm1}
\end{equation}
These equations are called equation of state (EOS) of the system.
\subsection{EVHRG Model}
In EVHRG Model, a short range hardcore repulsive hadron-hadron interaction is taken into account by considering excluded volume~\cite{EVHRG3,EVHRG4,EVHRG6} of the hadrons.

In EVHRG model pressure is given as
\begin{equation}
P(T,\mu_1,\mu_2,...)=P_{(m)}(T,\mu_1,\mu_2,...)+P_{(b)}(T,\mu_1,\mu_2,...)+P_{(\bar{b})}(T,\mu_1,\mu_2,...)
\end{equation}
\begin{equation*}
P_{(m)}(T,\mu_1,\mu_2,...)=\sum_{p}P_{(m)p}^{id}(T,\hat{\mu}_{(m)1},\hat{\mu}_{(m)2},...)
\end{equation*}
\begin{equation*}
P_{(b)}(T,\mu_1,\mu_2,...)=\sum_{q}P_{(b)q}^{id}(T,\hat{\mu}_{(b)1},\hat{\mu}_{(b)2},...)
\end{equation*}
\begin{equation*}
P_{(\bar{b})}(T,\mu_1,\mu_2,...)=\sum_{r}P_{(\bar{b})r}^{id}(T,\hat{\mu}_{(\bar{b})1},\hat{\mu}_{(\bar{b})2},...)
\end{equation*}
where $P_{(m)}$, $P_{(b)}$ and $P_{(\bar{b})}$ are mesonic, baryonic and anti-baryonic contribution to pressure. $\hat{\mu}_{(m)p}$, $\hat{\mu}_{(b)q}$ and $\hat{\mu}_{\bar{(b)}r}$ are the effective chemical potential for p-th meson, q-th baryon and r-th anti-baryons respectively which can be written as
\begin{equation*}
\hat{\mu}_{(m)p}=\mu_p-V_{ev,p}P_{(m)}(T,\mu_1,\mu_2,...)
\end{equation*}
\begin{equation*}
\hat{\mu}_{(b)q}=\mu_q-V_{ev,q}P_{(b)}(T,\mu_1,\mu_2,...)
\end{equation*}
\begin{equation}
\hat{\mu}_{(\bar{b})r}=\mu_r-V_{ev,r}P_{\bar{(b)}}(T,\mu_1,\mu_2,...)
\end{equation}
where $V_{ev,p} =4\frac{4}{3}\pi R_p^3$ is the excluded volume for p-th meson having hard-core radius $R_p$ and same treatment holds for baryons and anti-baryons.

Equations (5) and (6) are iteratively solved to get the pressure.
Since effective chemical potential $\hat{\mu}_{(m)p}$ is smaller than chemical potential $\mu_p$, pressure $P_{(m)}(T,\mu_1,\mu_2,...)$ is smaller than ideal mesonic pressure $P_{(m)}^{id}$ and same is true for baryons and anti-baryons.
From equations (5) and (6), we can calculate various thermodynamic quantities like number density of p-th meson $(n_{(m)p}^{E})$ as
\begin{equation}
n_{(m)p}^{E}=n_{(m)p}^{E}(T,\mu_1,\mu_2,...)=\frac{\partial P_{(m)}}{\partial\mu_p}=\frac{n_{(m)p}^{id}(T,\hat{\mu}_p)}{1+\sum_pV_{ev,p}n_{(m)p}^{id}(T,\hat{\mu}_k)}
\end{equation}
Similar relations hold for baryons and anti-baryons. In what follows we shall drop the notification $(m)$, $(b)$ and $(\bar{b})$ and the equations introduced will be valid for mesons, baryons and anti-baryons separately.

\subsection{Modified EVHRG Model}

As mentioned earlier, various authors have discussed a varying size
distribution for various hadron species~\cite{fluctuation1}.
To analyse the effect of unequal size of different hadron species, we
take recourse to virial expansion method. Such an analysis is done in
Ref.~\cite{st1,st2}. In this analysis, excluded volume for a single
particle is taken to be half the volume excluded by two touching spheres
of unequal radii instead of equal radii as taken in the previous EVHRG
analysis. We name this Model as Modified Excluded Volume Hadron
Resonance Gas Model (MEVHRG model).

The second virial coefficients of particle $"k"$ and $"n"$ are
\begin{equation}
a_{kn}=\frac{2}{3}\pi(R_k+R_n)^3=\frac{2}{3}\pi(R_k^3+3R_k^2R_n+3R_kR_n^2+R_n^3)
\end{equation}
Then pressure of the system becomes~\cite{st2}
\begin{equation}
P^M=T\sum_{k=1}^{N}\phi_ke^{\frac{\mu_k}{T}}\Big[1-\frac{4}{3}\pi R_k^3\sum_{n=1}^{N}\phi_ne^{\frac{\mu_n}{T}}-2\pi R_k^2\sum_{n=1}^{N}R_n\phi_ne^{\frac{\mu_n}{T}}-2\pi R_k\sum_{n=1}^{N}R_n^2\phi_ne^{\frac{\mu_n}{T}}\Big]
\end{equation}
where $\phi_k=g_k\gamma_S^{|S_K|}\int\frac{d^3p}{2\pi^3}\exp[-\frac{\sqrt{p^2+m^2}}{T}]$ is thermal density of particles and 'M' stands for MEVHRG model. Here $\gamma_S$ is the strangeness suppression factor and $|S_K|$ is the number of valence strange quarks and antiquarks in that particular hadron.
We make an approximation that surface and curvature terms in the above expression which are proportional to $R_k^2$ and $R_k$ respectively are equal and then we get
\begin{equation*}
P^M\simeq T\sum_{k=1}^{N}\phi_ke^{\frac{\mu_k}{T}}\Big[1-\frac{4}{3}\pi R_k^3\sum_{n=1}^{N}\phi_ne^{\frac{\mu_n}{T}}-4\pi R_k^2\sum_{n=1}^{N}R_n\phi_ne^{\frac{\mu_n}{T}}\Big]
\end{equation*}
To facilitate an iterative algorithm for obtaining $P^M$, we replace $\phi_ne^{\frac{\mu_n}{T}}\simeq\frac{P_n}{T}$
\begin{equation}
P^M\simeq T\sum_{k=1}^{N}\phi_ke^{\frac{\mu_k}{T}}\Big[1-\frac{4}{3}\pi R_k^3\frac{P^M}{T}-4\pi R_k^2\sum_{n=1}^{N}\frac{R_nP^M_n}{T}\Big]
\end{equation}
For $\mu_k/T<1$ we further approximate
\begin{equation}
P^M\simeq T\sum_{k=1}^{N}\phi_k\exp\Big[\frac{\mu_k}{T}-\frac{4}{3}\pi R_k^3\frac{P^M}{T}-4\pi R_k^2\sum_{n=1}^{N}\frac{R_nP^M_n}{T}\Big]
\end{equation}
Eqn. (11) can be written as
\begin{equation}
P^M=\sum_{k=1}^{N}P_k(T,\hat{\mu}_k^M)
\end{equation}
where
\begin{equation}
\hat{\mu}_k^M=\mu_k-\frac{4}{3}\pi R_k^3P^M-4\pi R_k^2\sum_{n=1}^{N}R_nP^M_n
\end{equation}
Eqn. (12) and (13) can be solved iteratively to get pressure.

Number density of i-th hadron is given as
\begin{equation}
n_i^M=\frac{(1+\sum_k4\pi R_k^3n_k^{id}(T,\hat{\mu}_k^M))n_i^{id}(T,\hat{\mu}_i^M)-\sum_j4\pi R_j^2n_j^{id}(T,\hat{\mu}_j^M)R_in_i^{id}(T,\hat{\mu}_i^M)}{\splitfrac{(1+\sum_k4\pi R_k^3n_k^{id}(T,\hat{\mu}_k^M))(1+\sum_j\frac{4}{3}\pi R_j^3n_j^{id}(T,\hat{\mu}_j^M))}{-\sum_j4\pi R_j^2n_j^{id}(T,\hat{\mu}_j^M)\sum_k\frac{4}{3}\pi R_k^4n_k^{id}(T,\hat{\mu}_k^M)}}
\end{equation}

\subsection{Effect of Lorentz Contraction (LMEVHRG Model)}
In a hadron gas, the constituting particles can have large kinetic energies and hence they can have large velocities. So the excluded volume of a particular particle in the rest frame of the thermal medium is Lorentz contracted.  This weakens the excluded volume repulsive interaction compared to the case where the effect of Lorentz contraction is not taken into account. Generalisation of cluster and virial expansions for momentum dependent inter-particle potentials to incorporate Lorentz contraction has been studied by some authors~\cite{st1}. The effect of Lorentz contraction has been introduced through a Lorentz contracted radius dependent potential. Calculating the Lorentz contracted excluded volume of two particles is very complex because the shape of the particles become ellipsoidal. Therefore, the distance between two touching ellipsoid depends not only on the size at the particles' rest frame but also on the angle of their relative velocities. The relevant area we are looking for is when the mean energy per particle is higher compared to the masses of the particles. The collisions with collinear velocities in the system is more common than collisions with non-collinear velocities since in the later case the effective excluded volume is larger and hence such collisions are suppressed. At reasonable densities, the orientation of different particles become correlated and hence the system tends to adjust itself in the configuration with minimum excluded volume. Also there is possibility of rotation of Lorentz contracted ellipsoids. Effect of rotation is negligible in a dilute system because the average number of collision per second among the particles of the system is very small. On the other hand, in a highly dense system the compact arrangement of particles hinders the ellipsoids from rotating. We shall ignore these effects in this work.  In this paper we take into account the effect of Lorentz contraction on the excluded volume of the particles in a simple phenomenological way in the rest frame of the system and compare the results with those of Lattice QCD. If a particle has momentum $p$, its size is contracted in the direction of motion according to the formula $R^{\prime}=R\sqrt{1-v^2}$ where v is the velocity of a particle. Now, $p=mv=\frac{m_0}{\sqrt{1-v^2}}v$ where $m_o$ is the rest mass of a particle. From this we get $1-v^2=\frac{m_0^2}{p^2+m_o^2}$

So, the effective chemical potential with Lorentz contraction becomes
\begin{equation}
\hat{\mu}_k^{LM}=\mu_k-\frac{4}{3}\pi R_k^3\frac{m
	_0}{\sqrt{p^2+m_0^2}}P^{LM}-4\pi R_k^2\frac{m
	_0}{\sqrt{p^2+m_0^2}}\sum_{n=1}^{N}R_nP^{LM}_n
\end{equation}
Here 'LM' in the superscript stands for Lorentz Contracted Modified EVHRG Model. Eqn. (15) is inserted into the momentum integral of eqn. (3) to calculate $P^{LM}$.

At a given temperature, the average velocity of lighter particles are larger than that of heavier particles and hence lighter particles experience larger Lorentz contraction than heavier particles. Lorentz contracted excluded volumes of different particles of the same species with different velocities are different and the two particle excluded volume matrix elements are different for different combinations of particles belonging to the same species but with different velocities. If in the rest frame of a particle species, the excluded volume of that species is larger then they will experience larger amount of volume contraction. The case is opposite for particle species with smaller rest volumes. If lighter particles are assigned larger excluded volumes than heavier particles then difference between the results of MEVHRG and LMEVHRG models will be greater than the case when lighter particles are assigned smaller excluded volumes.

There is a problem with this scenario we have considered to include the effect of Lorentz contraction. As the average excluded volume per particle decreases with increasing temperature, there is more available space in the system for new particles and hence phase transition is delayed when this model is used in combination with a model incorporating deconfinement. If one considers the effects of orientations and rotations of particles then this problem is circumvented up to some extent as they tend to decrease the particle density.

We shall now explore the effects of these various possible modifications
of the HRG model on different physical quantities. 

\section{Results} 
In this work we  have used baryon radius $R_b = 0.35 fm$, pion radius $R_\pi = 0.2 fm$ and radii of other mesons 
$R_m = 0.3 fm$. We have taken into account all particles listed in particle data book up to $3 \ GeV$ mass.

Analysis of hadron-hadron scattering in Ref.~\cite{Venugopalan:1992hy} indicates that there is little evidence for hardcore repulsive interaction in hadron pairs other than in nucleons. Generalisation of HRG model to excluded volume model with repulsive interaction among baryons only was considered in Ref.~\cite{Vovchenko:2016rkn}. There it was argued that baryon susceptibilities calculated within this framework are in good agreement with LQCD data. There is no clear evidence that short range repulsive excluded volume interaction exists among mesons as significant mesonic eigenvolume comparable to that of baryons leads to suppression of some thermodynamic functions and deteriorates the agreement with Lattice data~\cite{EVHRG7,Vovchenko:2014pka}. Interactions at short range, among baryon-antibaryon pairs, are found to be dominated by annihilation process. This consideration yielded satisfactory qualitative agreement of pressure and baryon susceptibilities with LQCD data~\cite{Vovchenko:2017xad}. However, as we have found in this work, it is difficult to contrast the LQCD data for electric charge susceptibilities with negligible mesonic excluded volume. Pions are the dominant contributor to electric charge susceptibilities and negligible pion volume overestimates LQCD data for electric charge susceptibilities. Here, we have considered meson-meson, baryon-baryon and antibaryon-antibaryon repulsive interactions only in this work. Inclusion of quantum correction to excluded volume requires knowledge of scattering phase shifts of different interaction channels which is 
an involved procedure. Such correction in EVHRG model was considered  in Refs.~\cite{Vovchenko:2017drx,typel}. In Ref.~\cite{Vovchenko:2017drx} the authors have obtained a monotonically decreasing second virial coefficient  with increasing temperature.  This suggests that effects of quantum correction decreases with the increase in temperature. We do not address such corrections in this work.

\subsection{Pressure}

In Fig 1, we have shown the scaled pressure ($P/T^4$) as a function of temperature for HRG, EVHRG, MEVHRG and LMEVHRG models and compared with Lattice data~\cite{Bazavov:2014pvz}.  The pressure corresponding to pure HRG is maximum. When we include hardcore repulsive interaction, there is a reduction in scaled pressure. Effect of repulsive interaction is prominent after $T=0.12 \ GeV$.

Scaled pressure for MEVHRG model is greater than that for EVHRG model since the effective chemical potential in MEVHRG case is larger than the EVHRG case. Scaled pressure for LMEVHRG model is even higher because in this case repulsive interaction is weaker. Scaled pressure for EVHRG and MEVHRG differ only slightly from each other. It can be argued that this difference will be greater if one considers different excluded volumes for different hadron species instead of only three types of excluded volumes. This effect can be expected from eqn. (13) where the last term in effective chemical potential deviates more from the EVHRG value as greater variety of radii is used. 

\begin{figure}[H]
	\begin{center}

		\includegraphics[scale=0.8]{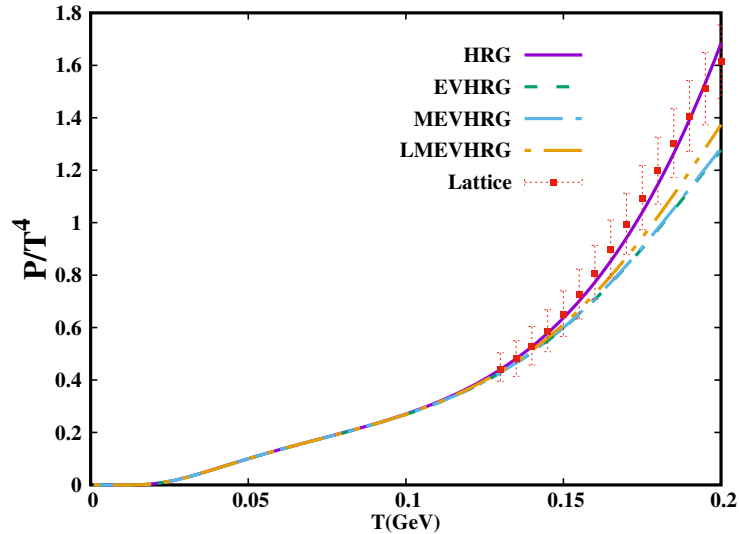}

	\label{fig 1}
	\caption{Variation of scaled pressure ($P/T^4$) with temperature.}
	\end{center}
\end{figure}

In this section we have found that the HRG model seems to give the best fit for pressure. However, we would like to look at other thermodynamic quantities also, like fluctuations, in the light of 
different variants of HRG which we will do in the next section.

\subsection{Fluctuations of Conserved Charges}
Fluctuations of conserved charges like net baryon number, electric charge, strangeness are useful indicators of thermalisation and hadronization of matter produced in ultra relativistic heavy ion collisions~\cite{a1,a2,a3}. Large fluctuations in various thermodynamic quantities are important signatures of the existence of Critical End Point (CEP) in the phase diagram.

Susceptibilities are defined as derivatives of grand canonical partition function $\mathcal{Z}$. The nth order susceptibility is defined as
\begin{equation}
\chi_n^x=\frac{1}{VT^3}\frac{\partial^n (ln\mathcal{Z})}{\partial (\frac{\mu_x}{T})^n}
\end{equation}
where $\mu_x$ is the chemical potential for conserved charge $x$, where $x$ may be B(baryon number), Q(electric charge) or S(strangeness). We calculate pressure at various chemical potentials around $\mu_x=0$ and find out the charge susceptibilities by fitting the data into Taylor expansion series.

We have shown in Figs. (2)-(3) second and fourth order susceptibilities of the conserved charges around zero chemical potentials ($\mu_B=\mu_Q=\mu_S$=0) and compare those with the LQCD results. From the plots one can see that susceptibilities for all the four models are almost same at low temperatures. As temperature increases, deviations among various models become prominent. For HRG model where no repulsive interaction is included, susceptibilities of all orders increase rapidly with temperature. Susceptibility for HRG Model is greater than corresponding EVHRG (where hardcore repulsion is included) case. Susceptibilities for MEVHRG model are little bit higher compared to EVHRG case for this choice of hadronic radii in this work. Whether the change between susceptibilities in EVHRG and MEVHRG model will be positive or negative depends on the choice of radii of hadrons and also on the type of conserved charge. If the dominating hadrons carrying a particular conserved charge are assigned larger excluded volume than that of dominated hadrons then susceptibility of that conserved charge in MEVHRG model will be lower than EVHRG case. This is because the last term in  eqn. (13) will be greater in MEVHRG model than its counterpart in EVHRG model. For the particular choice of hadronic radii in this work, all the susceptibilities in MEVHRG model are larger than corresponding EVHRG case. It is seen that $\chi_n^Q > \chi_n^S > \chi_n^B$. This is expected since susceptibilities at a certain temperature are governed by the dominating hadrons at that temperature carrying that charge. Since pions (lightest electrically charged hadrons) are lighter than kaons (lightest strange hadrons) which are lighter than protons (lightest baryons), hence the result.

\begin{figure}[h]
	\begin{center}

		\includegraphics[scale=0.4]{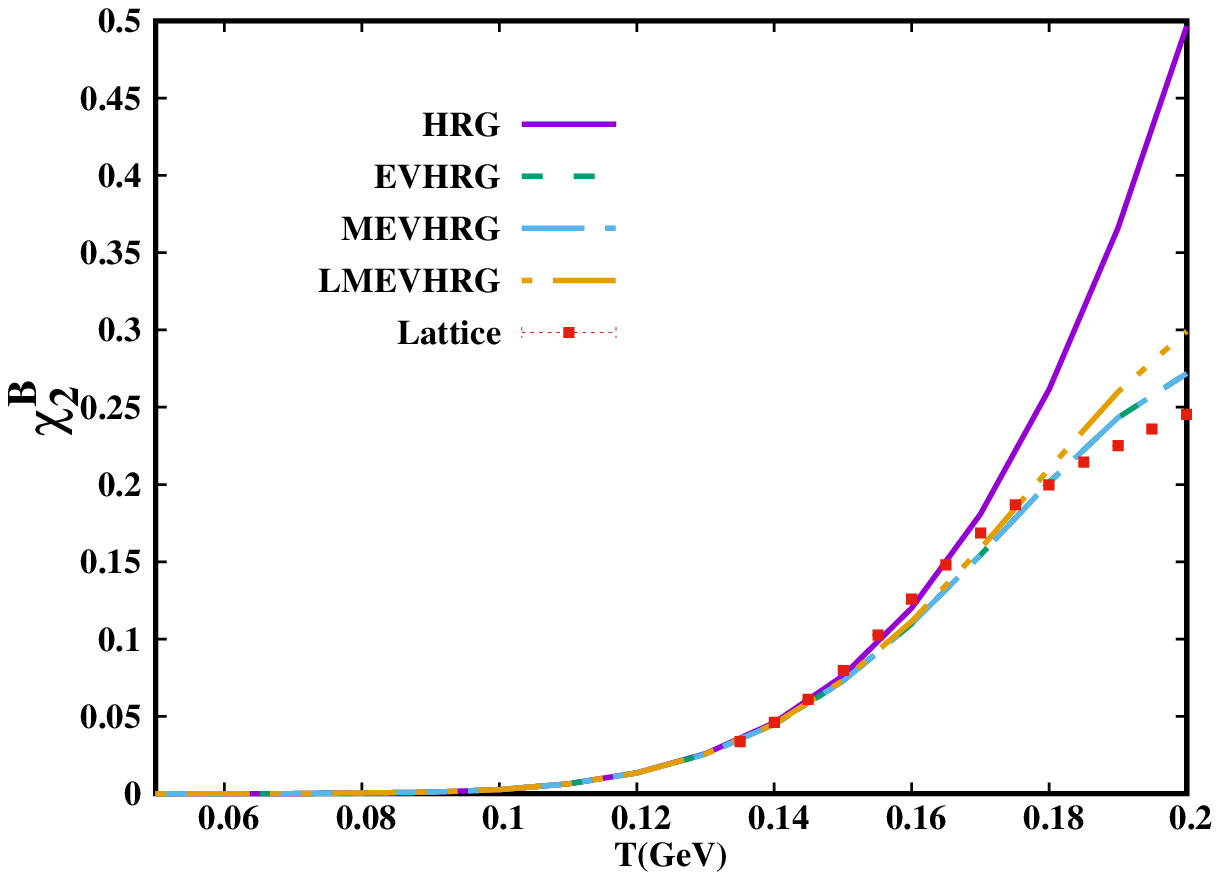}
		\includegraphics[scale=0.4]{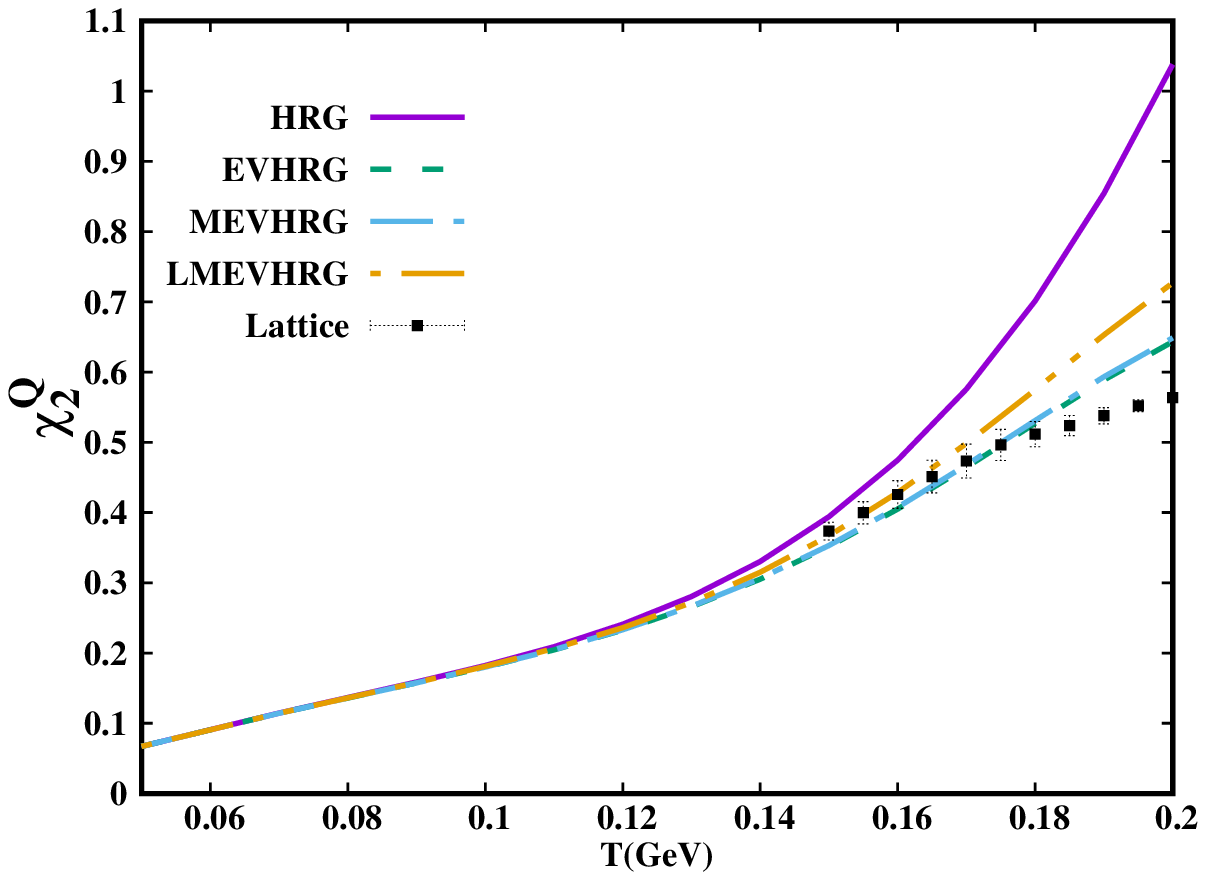}
		\includegraphics[scale=0.4]{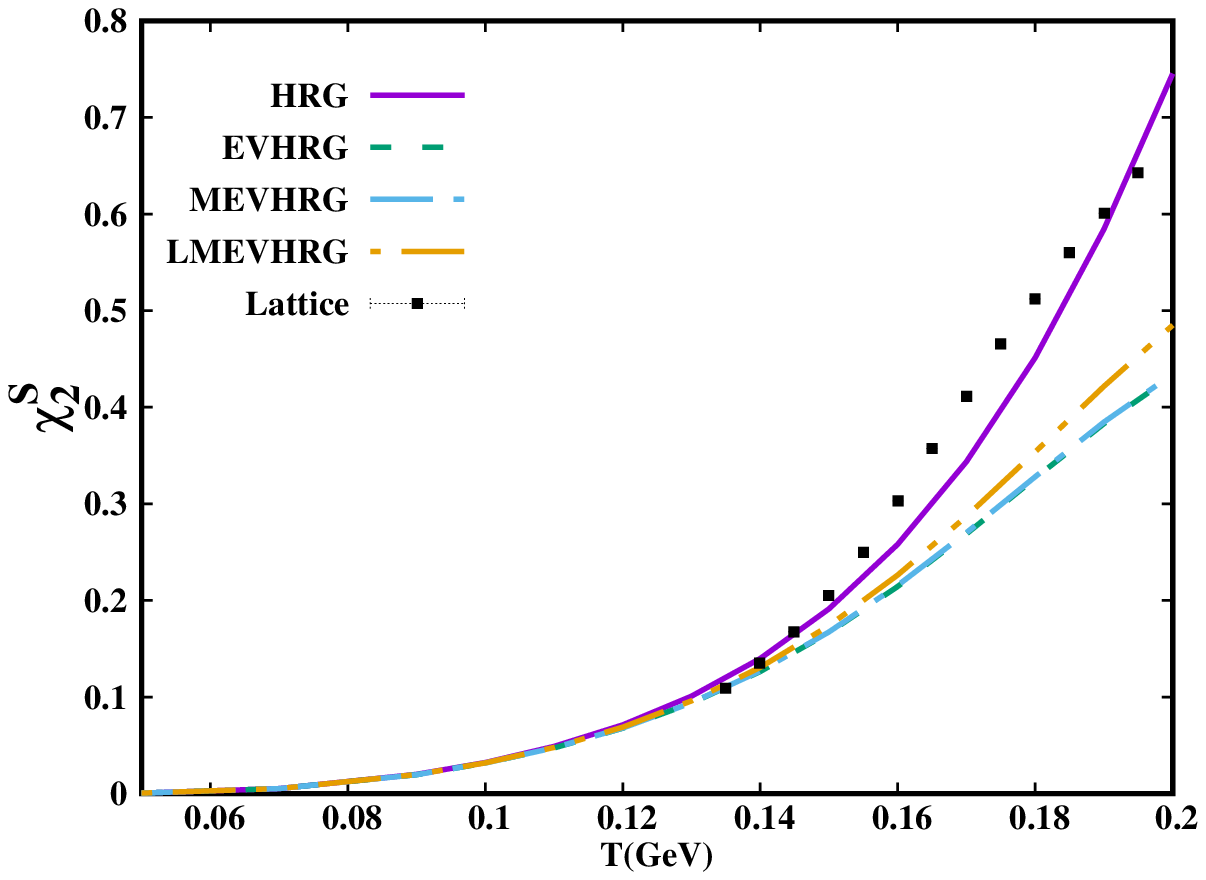}

	\caption{Second order susceptibilities for different conserved numbers}.
	\label{fig 2}
	\end{center}
\end{figure}

In Fig. \ref{fig 2} we have shown second order susceptibilities at zero chemical potentials ($\mu_B=\mu_Q=\mu_S$=0) 
for different conserved numbers and compare it with LQCD results given in Refs.~\cite{LQCD12,Borsanyi:2018grb}.

 In the top left panel we show $\chi_2^B$ as a function of temperature. 
It is seen that results of all the four models for 
$\chi_2^B$ coincide up to $T=0.14 \ GeV$. The EVHRG, MEVHRG and LMEVHRG results don't show significant difference up to somewhat higher temperature. Agreement with LQCD is quite satisfactory up to $T=0.18 \ GeV$ for all the three models with repulsive interaction.   This clearly shows the relevance of repulsive interaction. As expected, $\chi_2^B$ does not have any dependence on mesonic radii since there is no coupling among mesons and baryons in our model and mesons don't carry baryon number. Since the radii of all baryons are taken to be same, there is no quantitative difference between EVHRG and MEVHRG model results in this case. Difference between LMEVHRG model and MEVHRG model is small here and significant at high temperatures only as baryons need very high temperature to show significant Lorentz contraction.

In top right panel we have shown second order electric charge susceptibility.  It is seen that results of all the four models for $\chi_2^Q$ coincide up to $T=0.1 \ GeV$. EVHRG and MEVHRG results deviate from each other at medium range of temperature in contrast to $\chi_2^B$ as the dominant contributors, here, are mesons which have been assigned two different values of radii. Here the effect of Lorentz contraction is significantly larger than in $\chi_2^B$ as lighter mesons are the dominant contributors. The LMEVHRG data has the best match with LQCD data at medium range of temperatures but EVHRG and MEVHRG data give better agreement at somewhat higher temperatures. Once again 
we see that models with repulsive interactions perform better. 

In the bottom panel we have shown second order strangeness susceptibility. It is seen 
that results of all the four models, for $\chi_2^S$,
coincide up to $T=0.11 \ GeV$. One can see that $\chi_2^S$ deviates from LQCD data much more than that of
$\chi_2^B$ and $\chi_2^Q$. The lattice data overestimates even the pure HRG model results. Kaon is the most dominant contributor to
this susceptibility. Difference between MEVHRG and EVHRG results are insignificant here. Effect of Lorentz contraction here is smaller than that in $\chi_2^Q$ but stronger than that in $\chi_2^B$ as kaons are more massive than pions but lighter than nucleons.

\begin{figure}[H]
	\begin{center}

		\includegraphics[scale=0.4]{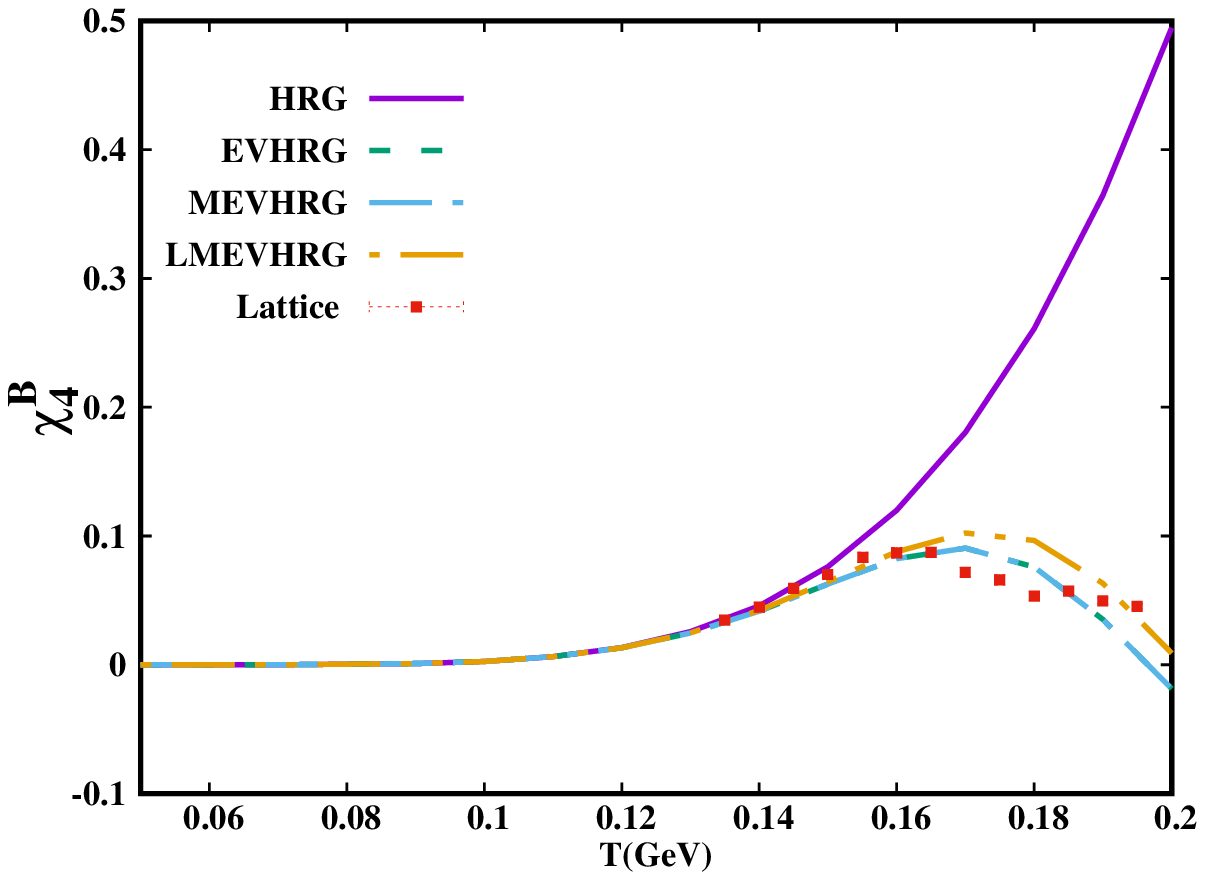}
		\includegraphics[scale=0.4]{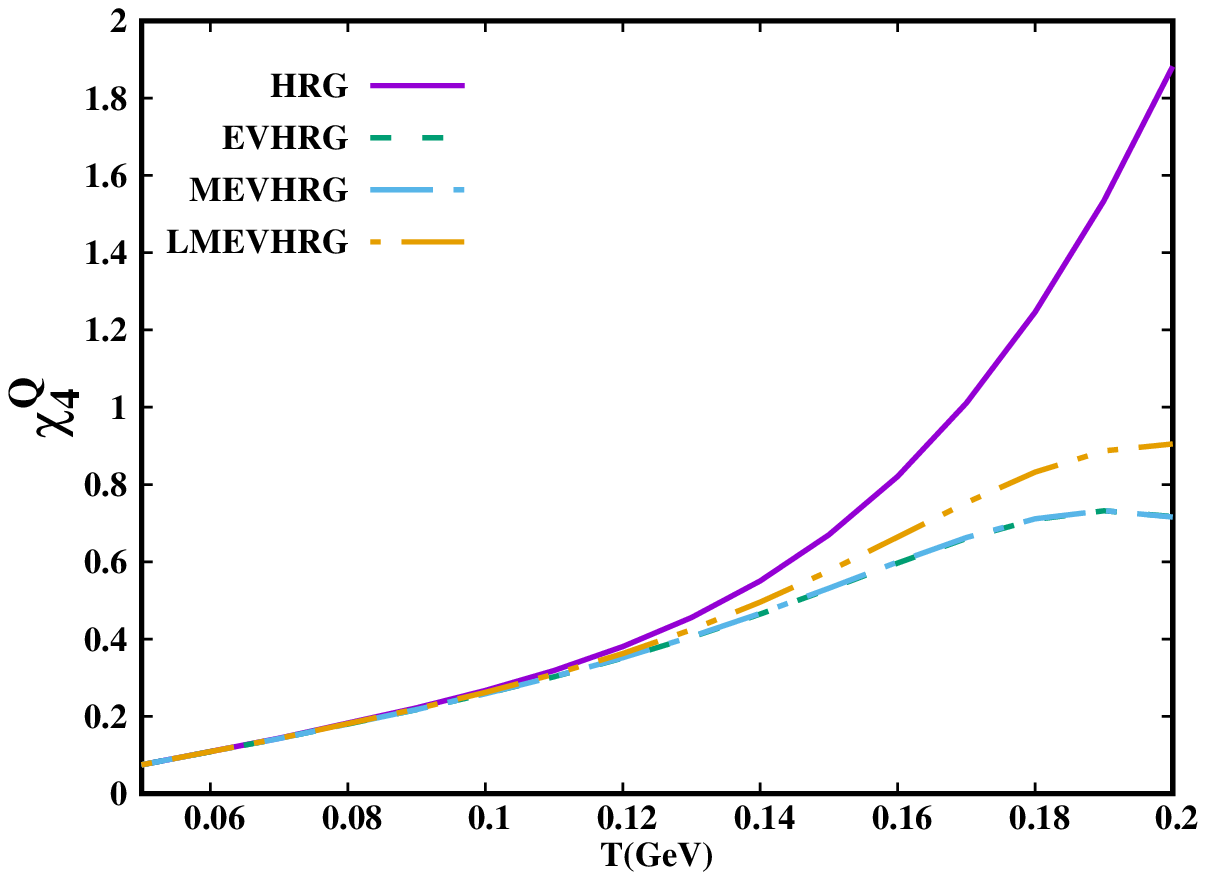}
		\includegraphics[scale=0.4]{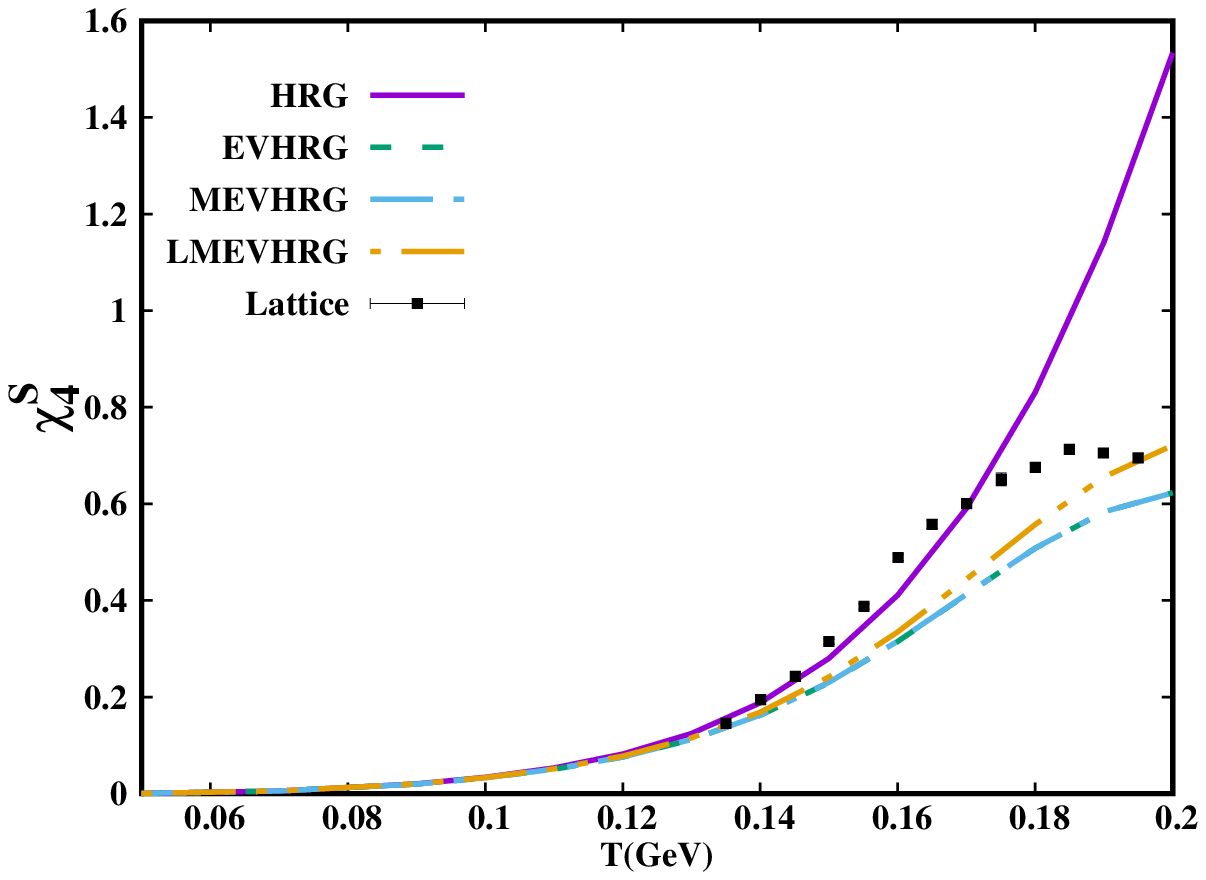}

	\caption{Fourth order susceptibilities for different conserved numbers}.
	\label{fig 3}
	\end{center}
\end{figure}

In Fig. \ref{fig 3} we have shown the fourth order susceptibilities of different conserved numbers at zero chemical potentials ($\mu_B=\mu_Q=\mu_S$=0) and compare those with LQCD results given in Ref.~\cite{Borsanyi:2018grb}. 

All the four model results for $\chi_4^B$, which is shown in the top left panel, coincide up to $T=0.12 \ GeV$. The results from different versions of interacting HRG model do not show significant difference up to somewhat higher temperature. For EVHRG, MEVHRG and LMEVHRG model, $\chi_4^B$ increases gradually with temperature, reaches a maximum and then starts to decrease. The results from the interacting models  have an excellent match with the LQCD data. Since we have taken only one type of baryonic radius, there is no difference in EVHRG and MEVHRG results in this case. Higher order cumulants of conserved charges are more sensitive to change in particle number and hence difference between MEVHRG and LMEVHRG model results for $\chi_4^B$ is larger than that for $\chi_2^B$ at high temperatures.

In the top right panel we have shown fourth order electric charge susceptibility. It is seen that, all the four model results, for $\chi_4^Q$, coincide up to $T=0.1 \ GeV$. The pion is the most significant contributor to $\chi_4^Q$. EVHRG and MEVHRG plots deviate negligibly from each other at all temperatures. The results for $\chi_4^Q$ shows a tendency of saturation at high temperatures. Like the baryon susceptibilities, $\chi_4^Q$ shows larger difference between MEVHRG and LMEVHRG model results than that of $\chi_2^Q$. Since we do not have 
proper lattice data for $\chi_4^Q$, with continuum extrapolation, we do not compare our results with any lattice data in this case.

In the bottom panel we have shown fourth order strangeness susceptibility.  The lattice data matches with HRG model results up to a temperature of $T=0.165 \ GeV$. Then it deviates from HRG results and approaches the interacting HRG results.

Though LQCD data for scaled pressure shows good agreement with HRG results,  baryon and electric charge susceptibilities significantly deviate from the HRG data. Even for the fourth order strangeness susceptibility the LQCD data deviates from the non-interacting scenario at high temperatures. On the other hand, the interacting models perform much better in reproducing the lattice results. This indicates that excluded volume correction is extremely relevant in describing the hot and dense strongly interacting matter.

Here we would like to mention that inclusion of Lorentz contraction gives a better fit with the lattice data in some cases. For $\chi_2^Q$, $\chi_2^B$ and $\chi_2^S$
the result with Lorentz contraction scenario has a maximum deviation of $10\%$ from the Lattice data and in some cases the deviation is less than $5\%$.

\section{Conclusion and discussion:}
In this work we have presented a modified version of EVHRG model where we have taken into account two additional things, namely, (1) effect of unequal radii for excluded volume of different particle species and (2) Lorentz contraction of excluded volume. Effect of unequal radii in EVHRG model has been studied in~\cite{Alba:2017bbr,Vovchenko:2016ebv} and effect of Lorentz contraction has been considered in~\cite{Oliinychenko:2012hj}. But these approaches are somewhat different than ours used here. We conclude the following important things:

(i) Though HRG model explains the scaled pressure, in order to explain baryon and electric charge susceptibilities repulsive models are absolutely essential. 

(ii) Strangeness susceptibility, of second order, in LQCD are higher than those in HRG. A similar result was obtained in Ref.~\cite{LQCD12}. The  
lattice data, for fourth order, matches with pure HRG result till $0.165 \ GeV$ of temperature. However, at higher temperature, it approaches the results with  
Lorentz contraction. It is argued that the strangeness sector gives incomplete picture about the particle spectrum and hence on thermodynamic quantities when only the hadronic states listed in Particle Data Group are considered~\cite{Alba:2017bbr,Alba:2017mqu}.

(iii) It is expected that a larger variety of excluded volume will shift the MEVHRG results in a larger quantity than EVHRG results. 

(iv) Inclusion of  Lorentz contraction has significant impact on pressure and susceptibilities as can be seen from our work and this is also important since LQCD inherently takes Lorentz contraction into account.

(v) Better agreement with LQCD data can be achieved by taking into account other modifications like temperature dependent particle mass instead of constant mass as used in this work and also some hybrid model with both HRG and quarks. We plan to address these in our future work.

\section{Acknowledgement}
The work is funded by  University Grants Commission (UGC) and Alexander von Humboldt (AvH) foundation and Federal Ministry
of Education and Research (Germany)  through Research Group Linkage programme.

\end{document}